\documentclass[11pt,a4paper]{article}
\RequirePackage[hyperref]{emnlp2020} 
\usepackage{times}
\usepackage{latexsym,amsmath,amsfonts,amssymb,bm}
\usepackage{microtype}

\RequirePackage{adjustbox}
\RequirePackage{soul}
\RequirePackage[normalem]{ulem} 

\RequirePackage{tikz,pgfplots}
\pgfplotsset{compat=1.14}
\RequirePackage[textsize=scriptsize]{todonotes}
\RequirePackage{enumitem}
\setlist[itemize]{nosep,leftmargin=*,labelwidth=0pt}
\setlist[enumerate]{nosep}
\setlist[description]{nosep,leftmargin=.8em}
\RequirePackage{colortbl}
\RequirePackage{nicefrac}
\RequirePackage[skip=1ex minus.5ex]{caption}

\intextsep \textfloatsep

\usepackage{multirow}

\makeatletter
\g@addto@macro{\normalsize}{%
\setlength{\abovedisplayskip}{0pt}%
\setlength{\abovedisplayshortskip}{0pt}%
\setlength{\belowdisplayskip}{0pt}%
\setlength{\belowdisplayshortskip}{0pt}}
\makeatother

\def\train{\text{tr}}
\def\KBtrain{\ensuremath{\text{KB}^\train}}
\def\KBpair{\ensuremath{\text{KB}^\text{Pair}}}
\def\dev{\text{de}}
\def\eval{\text{ev}}
\def\pred{\text{pr}}
\def\iou{\text{IOU}}
\def\giou{\text{gIOU}}
\def\aeiou{\text{aeIOU}}
\def\vol{\text{vol}}
\def\R{\mathbb{R}}
\def\C{\mathbb{C}}
\def\bs{\bm{s}}
\def\br{\bm{r}}
\def\bo{\bm{o}}
\def\bt{\bm{t}}
\def\bx{\bm{x}}
\def\Tall{\mathfrak{T}}
\def\rSO{\br^\text{SO}}
\def\rST{\br^\text{ST}}
\def\rOT{\br^\text{OT}}
\def\Normal{\mathcal{N}}

\def\sys{\textsc{TimePlex}}
\def\shortname{\sys}

\def\complex{CX}
\def\base{TX}

\aclfinalcopy
\def\aclpaperid{3617} 

\def\ztitle{Temporal Knowledge Base Completion: \\
New Algorithms and Evaluation Protocols}
\title{\ztitle}

\author{
Prachi Jain$^{\rm{*}1}$, 
{Sushant Rathi\thanks{\ \ \ Equal contribution}}$^{\ 1}$, 
Mausam$^1$ {\normalfont and} 
Soumen Chakrabarti$^2$
\\ 
$^1$ Indian Institute of Technology Delhi \\
$^2$ Indian Institute of Technology Bombay  \\
\{p6.jain, rathisushant5\}@gmail.com,
mausam@cse.iitd.ac.in,
soumen.chakrabarti@gmail.com
}

\begin{document}
\maketitle

\begin{abstract}
Research on temporal knowledge bases, which associate a relational fact $(s,r,o)$ with a validity time period (or time instant), is in its early days. Our work considers predicting missing entities (link prediction) and missing time intervals (time prediction) as joint Temporal Knowledge Base Completion (TKBC) tasks, and presents \sys, a novel TKBC method, in which entities, relations and, time are all embedded in a uniform, compatible space. \sys\ exploits the recurrent nature of 
some facts/events 
and temporal interactions between pairs of relations, yielding state-of-the-art results on both prediction tasks.

We also find that existing TKBC models heavily overestimate link prediction performance due to imperfect evaluation mechanisms. In response, we propose improved TKBC evaluation protocols for both link and time prediction tasks, dealing with subtle issues that arise from the partial overlap of time intervals in gold instances and system predictions.
\end{abstract}

\section{Introduction}
\label{sec:Intro}

A knowledge base (KB) is a collection of triples $(s,r,o)$, with a subject $s$, a relation type $r$ and an object $o$. KBs are usually incomplete, necessitating completion (KBC), i.e., inferring facts not provided in the KB. A KBC model is often evaluated via link prediction: supplying missing arguments to queries of the form $(s,r,?)$ and $(?,r,o)$.

Many relations are transient or impermanent.  
Temporal KBs annotate each fact (event) with the time period (instant) in which it holds (occurs) \cite{yago2-ijcai-HoffartSBW13}. A person is born in a city in an instant, a politician can be a country's president for several years, and a marriage may last between years and decades. Temporal KBs represent these by $(s,r,o,T)$ tuples, where $T$ is a span of time. Temporal KBC (TKBC) performs completion of temporal KBs. It is also primarily evaluated by link prediction queries $(s,r,?,T)$ and $(?,r,o,T)$. Recently, time prediction $(s,r,o,?)$ has also been considered for predicting time instants, but not time intervals \cite{lacroix2020tensor}.


While KBC has been intensely researched, TKBC is only beginning to be explored. TKBC presents novel challenges in task definition and modeling. For instance, little is known about how best to predict intervals for $(s,r,o,?)$ queries, or how to evaluate a system response interval. Moreover, we show that even for link prediction queries, evaluation faces subtle complications owing to the inclusion of $T$ in $(s,r,?,T)$ queries and requires careful rethinking of evaluation protocols. In this paper, we propose improved evaluation protocols for both link and time prediction tasks in a TKBC.

TKBC also brings unique modeling opportunities. A TKBC system can learn typical durations of relation validity, or distributions over time gaps between events, from training data. E.g., a person must be born before becoming president, which must precede death. A nation rarely has two presidents at the same time. Such constraints can better inform both link and time predictions.

In response, we present \shortname, a novel TKBC model, which obtains state-of-the-art results 
on benchmark datasets for both link and time prediction. At a high level, \shortname{} performs tensor factorization of a temporal KB, using complex-valued embeddings for relations, entities and time points. It enables these embeddings to capture implicit temporal relationships across facts and relations, by providing temporal differences as explicit features.
Our contributions are summarized as:
\begin{itemize}[leftmargin=*,labelwidth=0pt]
\item We propose evaluation protocols for link and time interval prediction queries for TKBC. For link prediction, we highlight that existing evaluations seriously over/under-estimate system performance, and offer a time-aware filtering method for more reliable evaluation. For time interval prediction, we propose an evaluation metric that rewards a model for predicting an interval with partial overlap with gold interval, as well as for nearness to gold in case of no overlap.

\item We present \sys, a TKBC model that factorizes a temporal KB using entity, relation and time embeddings. It can learn and exploit soft ordering and span constraints between potentially all relation pairs (including that of a relation with itself). It beats recent and competitive models on several recent standard TKBC data sets.
\end{itemize}

\noindent We will release an open-source implementation\footnote{github.com/dair-iitd/tkbi} of all models and experiments discussed here.

\section{Preliminaries and Prior Work}
\label{sec:Prelim}

\subsection{Time-Agnostic KBC}
Time-agnostic KBC has been intensely researched
\citep{BordesUGWY2013TransE, Yang+2015DistMult, NickelRP+2016holographic, DBLP:conf/acl/JainKMC18,LacroixUO2018canonical, JainMMC2018MfTfOov}. A common approach is to 
score an $(s,r,o)$ triple as a function over jointly learned entity and relation embeddings. The models are trained using loss functions imposing - scores for known triples should be higher than (randomly sampled) negative triples.

Our work is based on ComplEx \citep{TrouillonWRGB2016Complex}, abbreviated as \complex{}. It embeds $s,r,o$ to vectors of complex space $\bs, \br, \bo \in \C^D$. \complex{} defines the score $\phi$ of a fact $(s,r,o)$ as $\operatorname{Re}(\langle \bs, \br, \bo^\star\rangle)$ where
\begin{align}
\langle \bs, \br, \bo^\star \rangle &= \textstyle
\sum_{d=1}^D s[d] \; r[d] \; o^\star[d] \label{eq:DMCX}
\end{align}
is a 3-way inner product, $\bo^\star$ is the complex conjugate of~$\bo$, and $\operatorname{Re}(c)$ is real part of~$c\in\C$. If real embeddings are used instead, the above formula reduces to DistMult~\citep{Yang+2015DistMult}.
We choose \complex{} as our base model, because it 
is competitive with recent KBC models \citep{Ruffinelli2020You}.



\subsection{Temporal KBC Problem Setup}
\label{sec:Prelim:TempSetup}

A temporal KB associates the validity of a triple $(s,r,o)$ with one or more time intervals $T \subseteq \Tall$, where $\Tall$ is the domain of ``all time". Each interval $T$ is represented as $[t_b,t_e]$, with begin and end time instants. Some event-style facts (e.g., born in) may have $t_b=t_e$. For simplicity, we assume that $\Tall$ is discretized to a suitable granularity and is represented by a set of integers. Temporal KB facts have the form $(s,r,o,T)$, and are partitioned into train, dev and eval (test) folds, abbreviated as $\train, \dev, \eval$. System predictions are abbreviated as~$\pred$. 

Given the train and dev folds, our goal is to learn a model that scores any unseen fact. A system is evaluated via link prediction queries $(?,r,o,T)$ and $(s,r,?,T)$, and time interval prediction queries $(s,r,o,?)$. In our setting, KB incompleteness exists at all times --- the eval fold may include instances from any interval in time, arbitrarily overlapping train and dev fold instances.\footnote{A different TKBC task studies only future fact predictions \citep{Trivedi+2017KnowEvolve, Jin+2019ReNet}.}

\subsection{Recent TKBC Systems}

Recent work adopts a common style for extending $\phi(s,r,o)$ to temporal score $\phi(s,r,o,t)$. \citet{lacroix2020tensor} embed each time instant $t$ to vector $\bt$ and use the form $\langle \bs, \br, \bo^\star, \bt\rangle$ (called TNT-ComplEx). This can be interpreted as any \emph{one} of $\bs,\br,\bo^\star$ becoming $\bt$-dependent.
\citet{goel2019diachronic} make \emph{both} subject and object embeddings time-dependent; the `diachronic' embedding $\bm{e} \in \R^D$ of entity $e$ is characterized by $\bm{e}_t[d] = a_e[d]\sin(w_e[d]\,t + b_e[d])$, where $d\in D$ and the sinusoidal nonlinearity affords the capacity to switch ``entity features'' on and off with time~$t$. HyTE \cite{DasguptaRT2018HyTE} models $\bt\in\R^D, \|\bt\|_2=1$ and project \emph{all} of $\bs,\br,\bo$ on to $\bt$:
$\bx\downarrow\bt = \bx - (\bx \cdot \bt) \bt, \; \text{where} \;
\bx\in\{\bs,\br,\bo\}$. In all cases, time-dependent entity embeddings are plugged into standard scoring functions like DistMult, CX, or SimplE \citep{kazemi2018simple}. A very different approach \cite{Garcia+2018TADM} encodes the string representation of relation and time with an LSTM, which is used in TransE (TA-TransE) or DistMult (TA-DM). 

These formulations do not directly model recurrences of a relation or interactions (e.g., mutual exclusion) between relations. There is some prior work on explicitly providing ordering constraints between relations (e.g., born, married, died) \citep{Jiang+2016TkbcILP}. In contrast, \shortname{} assumes no such additional engineered inputs; it has explicit components to enable learning of temporal (soft) constraints, as model weights, jointly with embeddings of entities, relations, and time instants. Such constraint based reasoning has also been exploited (in a limited way) for a different task, namely, temporal question answering~\cite{tqa-cikm-JiaARSW18}.

\subsection{Standard Evaluation Schemes}
\paragraph{Link Prediction:} Link prediction queries in KBC are of the form $(s,r,?)$ with a gold response~$o^\eval$. Similarly, for TKBC they are of the form $(s,r,?,T)$. The cases of $(?,r,o)$ and $(?,r,o,T)$ are symmetric and receive analogous treatment. Link prediction performance is evaluated by finding the rank of $o^\eval$ in the list of all entities ordered by decreasing score $\phi$ assigned by the model, and computing MRR. Other measures include the fraction of queries where $o^\eval$ is recalled within the top 1 or top 10 ranked predictions (HITS@1 and HITS@10). 

A query may have multiple correct answers. A model must not be penalized for ranking a different \emph{correct} entity over $o^\eval$. In KBC this is achieved by filtering out all correct entities above $o^\eval$ in ranked list before computing the metrics. In TKBC, filtering requires additional care, as depicted in Table~\ref{table:filtering}. We develop time-aware filtering in Section~\ref{sec:FilterEval:LinkPred}.


\paragraph{Time Prediction:} Time prediction queries of the form $(s,r,o,?)$ will require comparing a gold time interval $T^\eval = [t_b^\eval, t_e^\eval]$ with a predicted interval $T^\pred = [t_b^\pred, t_e^\pred]$. Since this is an understudied task, evaluation metrics have not yet been standardized. One might adapt the TAC metric popular in Temporal Slot Filling \citep{JiGDG2011TAC, Surdeanu2013TAC}. Adapted to TKBC, TAC\footnote{TAC's original score compares gold and predicted \emph{bounds} on begin and end of an interval. This formula is its adaptation, where begin and end are each a specific time point.} will compute a score as $\frac{1}{2}\!\left[
\frac{1}{1+|t_b^\eval-t_b^\pred|} +
\frac{1}{1+|t_e^\eval-t_e^\pred|} 
 \right]$.
Unfortunately, TAC score is not entirely satisfactory for this task. For instance, TAC will assign the same merit score when gold interval [10,20] is compared with predicted interval [5,15], versus when gold [100,200] is compared with prediction [95,195]. However, a human would judge the latter more favorably, because a 5-minute delay in a 10-minute trip would usually be considered more serious than in a 100-minute journey. In response, we investigate alternative evaluation metrics inspired by bounding box evaluation protocols from Computer Vision, in Section~\ref{sec:FilterEval:TimePred}.

\begin{table*}
\centering
\begin{small}
\begin{tabular}{|c|c||c||c||c|c|c|c|} \hline
\multicolumn{8}{|c|}{\textbf{Eval query:}
$(s=\text{French National Assembly}, r=\text{has member}, o=?, T^\eval=[2000, 2003])$
} 
\\ \hline
\rowcolor{gray!10}
Candidates & Known
& \textbf{Method~1} & \textbf{Method~2} &
\multicolumn{4}{c|}{\textbf{Method~3}} \\
\rowcolor{gray!10}
$o$,~system & duration of & Unfiltered & Time- &
\multicolumn{4}{c|}{Time-sensitive} \\
\rowcolor{gray!10}
ordered & $o$~(any fold) & & insensitive & 2000 & 2001 & 2002 & 2003 \\ \hline
Pierre  & $[2002, 2003]$ & 1 & 0  & 1 & 1 & 0 & 0\\ 
Paul    & $[2003,2008]$ & 1 & 0 & 1 & 1 & 1 & 0\\ 
Alain & $[2008,2009]$ & 1  & 0   & 1 & 1 & 1 & 1\\ 
Claude & $[2000,2003]$ & 1  & 0 & 0 & 0 & 0 & 0\\ 
\emph{Jean} & -  & - & - & - & - & - & -\\ \hline
\rowcolor{gray!10}
\multicolumn{2}{|c|}{Time-sensitive rank of \emph{Jean}}
& 1+4=5 & 1+0=1 & 1+3=4 & 1+3=4 & 1+2=3 & 1+1=2\\ \hline
\end{tabular}
\caption{\emph{Jean} is the gold answer ($o^\eval$). Rows are ranked system predictions, which may be seen with same $s$ and $r$ for different intervals (Column 2). Columns 3--4 show the filtering of existing methods (1:unfiltered, 0:filtered).
Columns 5--8 (Method~3, our proposal) show the filtering for each time instant. The bottom row shows ranks of \emph{Jean} as computed by different methods. Existing methods over- or under-estimate performance. Method~3 assigns \emph{Jean} a rank of 3.25, which is the average of the filtered ranks $\{4,4,3,2\}$ for each time instant in $[2000,2003]$.}
\label{table:filtering}
\end{small}
\end{table*}

\section{Evaluation Metrics and Filtering}
\label{sec:FilterEval}


The preceding discussion motivates why we need clearly-thought-out filtering and evaluation schemes, not only for time interval prediction queries, but also because time affects link prediction evaluation in subtle but fundamental ways. This section addresses both issues.

\subsection{Time Interval Prediction}
\label{sec:FilterEval:TimePred}


One possible way to evaluate time prediction is to adapt measures to compare bounding boxes in computer vision, e.g., Intersection Over Union (IOU): $\iou(T^\eval,T^\pred) = \frac{\vol(T^\eval\cap T^\pred)}{\vol(T^\eval\cup T^\pred)}\in[0,1]$, where $\vol$ for our case simply refers to the size of the interval. Unfortunately, $\iou$ loses discrimination once $T^\eval\cap T^\pred=\varnothing$; e.g., $\iou([1,2],[3,4])=\iou([1,2],[30,40])=0$. This has been noticed recently in computer vision also, and a metric called gIOU been introduced \citep{RezatofighiTGSRS2019gIOU}:  
\begin{multline}
\giou(T^\eval,T^\pred) = \iou(T^\eval,T^\pred) - \\
\frac{\vol((T^\eval\Cup T^\pred)\setminus
(T^\eval\cup T^\pred))}{\vol(T^\eval \Cup T^\pred)} \in (-1,1].
\label{eq:gIOU}
\end{multline}
$T^\eval\Cup T^\pred$ is the smallest single contiguous interval (\textbf{hull}) containing all of $T^\eval$ and $T^\pred$.  E.g., $[1,2]\Cup[30,40]=[1,40]$. 

$\giou$ can be negative, which is not ideal for a performance metric that is aggregated over instances. A simple fix ($\giou'$) is to scale it to [0,1] via $(\giou+1)/2$, but we notice that the tiniest overlap between $T^\eval$ and $T^\pred$ yields $\giou'$ to be at least half, regardless of $\vol(T^\eval)$ or $\vol(T^\pred)$. In response, we propose a novel \emph{affinity enhanced $\iou$}: 
\begin{align}
\aeiou(T^\eval,T^\pred) &\!=\!
\frac{\max\{1, \vol(T^\eval\cap T^\pred)\}}
{\vol(T^\eval\Cup T^\pred)}
\label{eq:aeiou}
\end{align}
When $T^\eval\cap T^\pred=\varnothing$, the denominator includes ``wasted time'', reducing $\aeiou$.  The `$1$' in the numerator represents the smallest granularity of time in the data (see Section~\ref{sec:Prelim:TempSetup}).

\paragraph{Comparison of Evaluation Metrics: } 
A good time interval prediction metric ($M$) must satisfy the property ($P$) that: if two predicted intervals have intersections of the same size (possibly zero) with the gold interval, then the prediction that has a smaller hull with the gold interval should be scored higher by~$M$.
Formally, let $T^{\pred_1}$ and $T^{\pred_2}$ be two predictions made for~$T^{\eval}$. 

\noindent{\bf Property~P:} Let $\vol(T^\eval \cap T^{\pred_1})= \vol(T^\eval \cap T^{\pred_2})$. Then, $M(T^{\eval},T^{\pred_1})>M(T^{\eval},T^{\pred_2})$ if and only if $\vol(T^\eval \Cup T^{\pred_1})< \vol(T^\eval \Cup T^{\pred_2})$.




\noindent{\bf Theorem:} $\iou$ and $\giou'$ do not satisfy property~P, whereas $\aeiou$ satisfies it. 

The proof for the theorem is in Appendix \ref{sec:app:DiscussionTimeMetrics}. This suggests that $\aeiou$ is a more defensible metric for our task, compared to other alternatives. 

\subsection{Link Prediction}
\label{sec:FilterEval:LinkPred}

We first illustrate the unique challenges offered by TKBC link prediction queries through an example in Table \ref{table:filtering}. 
The query asks for the name of a person who was a member of the French National Assembly in interval $[2000,2003]$. 
Let the gold answer (object) $o^\eval$ be Jean, which is ranked at the fifth position by the model. All four entities above Jean are seen with the same subject and relation in the data, but for different time intervals. E.g., Pierre is also a member of the assembly, but during $[2002,2003]$. The key question is: how should the four entities above Jean be filtered to compute its final rank?\\
We argue (Table~\ref{table:filtering}) that existing filtering approaches are unsatisfactory. \citet{DasguptaRT2018HyTE} underrate model performance by not performing any filtering (Method~1). In this example, the model is penalized for Claude, even though the time-interval for Claude exactly matches the query. On the other hand, \citet{Garcia+2018TADM} and \citet{Jin+2019ReNet} ignore time information altogether and filter out \emph{all} entities seen with gold $(s, r)$. This can greatly overestimate system quality (Method~2). For instance, the model is not penalized for predicting Alain, even though its membership interval has no overlap with the query interval. \\
Ideally, filtering must account for the overlap between the query time interval and the time intervals associated with system-proposed entities. We propose such a filtering strategy (Method~3). We split the query interval into time instants, and compute a filtered rank for each time point independently. Entities that have full time overlap (or no overlap) will always (respectively, never) get filtered for a time instant. Partially overlapping entities will get filtered in only overlapping instants (e.g., 2 out of 4 for Pierre). 
After computing filtered ranks for each time instant, we output the final rank as an average of all such filtered ranks. In this example, this approach will compute the average of $\{4,4,3,2\}$, which is~$3.25$. This average rank is used when computing standard metrics like MRR and HITS@10.

Note that the run-time complexity of the proposed evaluation protocol is linear in the size of interval, because we compute a filtered rank for each time point separately.

\section{The Proposed \shortname\ Framework}
\label{sec:TimePlex}

Similar to TNT-Complex, \shortname{} learns complex-valued entity, relation, and time instant embeddings. However, it has several differences from TNT-Complex. (1)~Its base scoring function $\phi^{TX}(s,r,o,t)$ adds several products of three embeddings, instead of a single four-way product (Section~\ref{sec:TimePlex:Score}). (2)~It has a fully automatic mechanism to introduce additional features to capture recurrent nature of a relation, as well as temporal interactions between pairs of relations (Section~\ref{sec:TimePlex:Pair}). (3)~It uses a two-phase training (Section~\ref{sec:TimePlex:Train}) curriculum that estimates first the embeddings and then novel additional parameters. (4)~Its testing protocol can output a missing time-interval $T$ for time-interval prediction queries (Section~\ref{sec:TimePlex:Test}).




\subsection{\shortname{} Base Model}
\label{sec:TimePlex:Score}

Just as a joint distribution is often approximated using lower-order marginals in graphical models \citep{KollerF2009GraphicalModels}, \shortname{} constructs a base score ($\phi^{\base{}}$) by augmenting CX score with three time-dependent terms:\\ $\phi^{\base{}}(s,r,o,t)$
\begin{multline}
= \langle \bs, \rSO, \bo^\star\rangle + 
\alpha\,\langle \bs, \rST, \bt^\star\rangle \\
+ \beta\, \langle \bo, \rOT, \bt^\star\rangle
+ \gamma\,\langle \bs, \bo, \bt^\star\rangle.
\label{eq:srot}
\end{multline}
Here, $\bm{s},\bm{o},\bm{t}\in\C^D$, whereas each $r$ is represented as a collection of three such vectors $(\rSO,\rST,\rOT)$, and hence requires three times the parameters. $\rST$ represents a relation which is true for entity $s$ at time $t$ (similarly for $\rSO$ and $\rOT$). $\alpha$, $\beta$ and $\gamma$ are hyperparameters.

\citet{Jiang+2016TkbcILP} observed that several relations attach to a subject or object only at specific time points. E.g., subject Barack Obama was president in 2009, regardless of the object United States. In such cases, the formulation above is fully expressive.
To extend from single time instants $t$ to an interval $T$, we propose
\begin{align}
\phi^{TX}(s,r,o,T) &= \textstyle \sum_{t\in T} \phi^{TX}(s,r,o,t).
\end{align}

\subsection{Relation Recurrence and Pair Scores}
\label{sec:TimePlex:Pair}

We extend \shortname's base model via additional (soft) temporal constraints that can help in better assessing the validity of a tuple. We aim to capture three types of temporal constraints: 
\begin{description}
\item[Relation Recurrence:] Many relations do not recur for a given entity (e.g., a person is born only once). Some relations recur with fixed periodicity (e.g., Olympic games recur every four years). Recurrences of other relations may be distributed around a mean time period.
\item[Ordering Between Relations:] A relation precedes another, for a given entity. E.g., 
\emph{personBornYear} should precede \emph{personDiedYear} for a given subject entity (person).
\item[Time Gaps Between Relations:] The difference in time instants of two relations (wrt to an entity) is distributed around a mean, e.g., \emph{personDiedYear} minus \emph{personBornYear} has a mean of about 70 with some observed variance.
\end{description}
\noindent
The first constraint concerns a single relation, whereas the latter two concern pairs of relations. \citet{Jiang+2016TkbcILP} attempted to capture relation ordering constraints as model regularization, but their approach does not take into account time differences. Nor does it model relation recurrence.

Basic \shortname{} may not be able to learn these constraints from data either, since each time instant is modeled as a separate embedding with \emph{independent} parameters --- it has no explicit understanding of the difference between two time instants. 
In response, we augment \shortname{} with additional features that capture how soon an event recurs, or how soon after the occurrence of one relation, another relation is likely to follow. We define two scoring functions $\phi^{\text{Rec}}$ and $\phi^{\text{Pair}}$ for these two cases, to be aggregated with $\phi^{\base{}}$~(eqn.~\ref{eq:srot}).

\begin{figure*}
\centering
\begin{tabular}{|cc|}
\hline 
(a) &
\includegraphics[width=1.3\columnwidth]{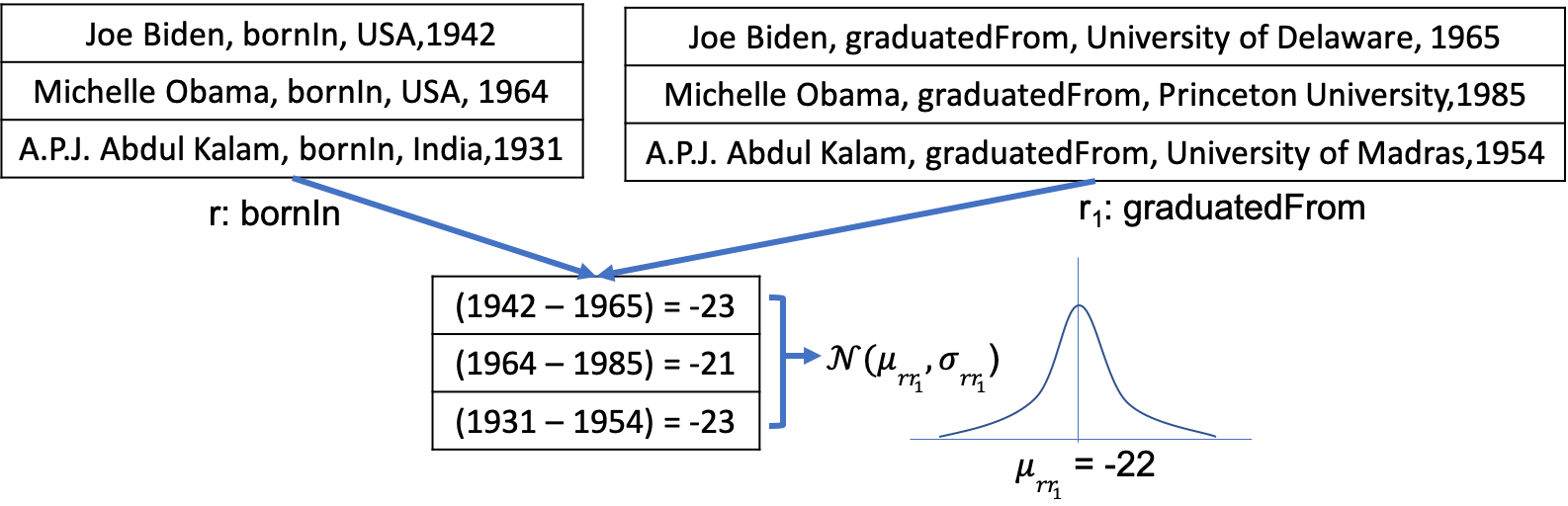} \\ \hline
(b) &
\includegraphics[width=1.3\columnwidth]{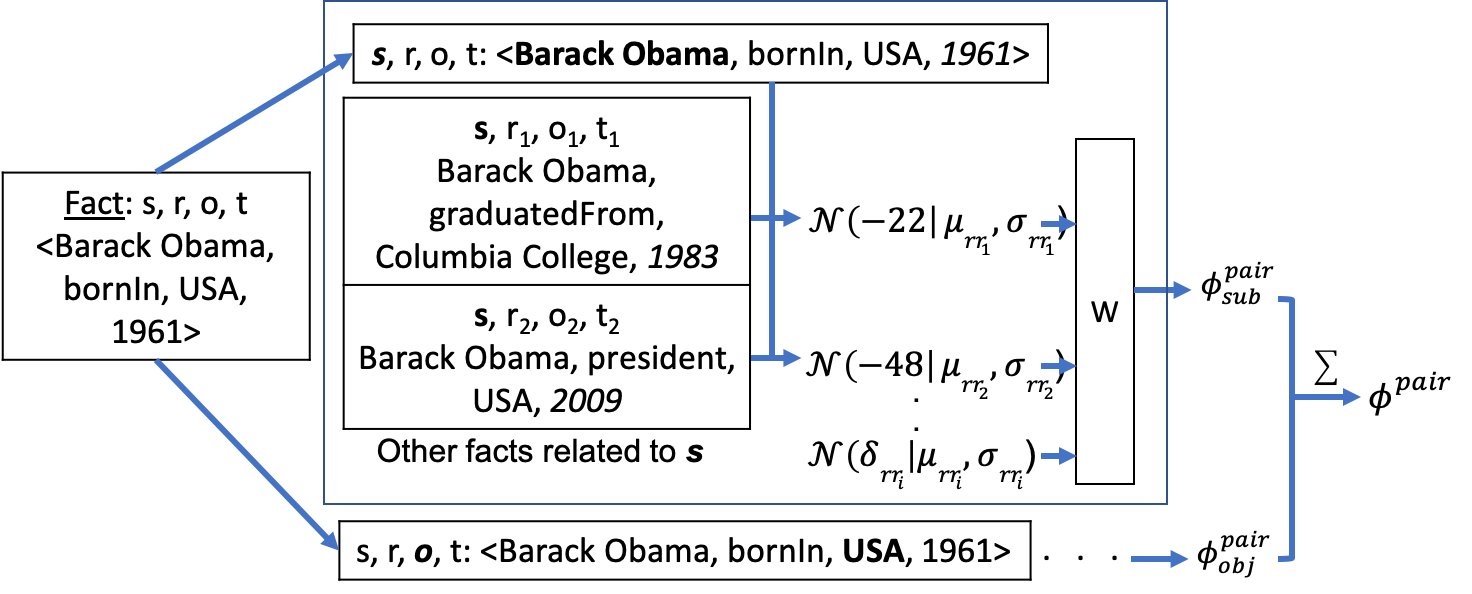}
\\ \hline
\end{tabular}
\caption{(a)~{\it Pre-training Data Statistics Collection Strategy} for relation pair (bornIn, graduatedFrom). Such statistics are computed for all relation pairs, and (b)~{\it Relation Pair Score Computation} of a fact using the statistics collected in part~(a). Here, $\delta_{rr_i} = (t - t_i)$.}
\label{fig:block-dig}
\end{figure*}

Inspired by \citet{GarciaduranN2018KbLRN},
we model time gaps as drawn from Gaussian distributions. We use $\Normal(x|\mu,\sigma)$ to denote the probability density of a Gaussian distribution with mean $\mu$ and std deviation $\sigma$ at the time (difference) value~$x$ (See Figure~\ref{fig:block-dig}\,(a)). 
We denote as $\KBtrain$ all tuples in the train fold.

\paragraph{Recurrence Score:} 
We say that $(s,r,o)$ \emph{recurs} if there are at least two distinct intervals $T$ such that $(s,r,o,T)\in\KBtrain$.
If there are at least $K^{\text{Rec}}$ distinct pairs $(s,o)$ such that $(s,r,o)$ recurs, then $r$ is considered \textit{recurrent}.
$K^{\text{Rec}}$ is a hyperparameter.

For each recurrent relation $r$, our model learns three new parameters: $\mu_r$, $\sigma_r$, and $b_r$. Intuitively, $\Normal(\cdot|\mu_r,\sigma_r)$ represents a distribution of typical durations between two recurring instances of a relation (with a specific subject and object entity) and $b_r$ is the bias term. For non-recurrent relations, only the bias $b_r$ is learnt.
While computing recurrence features, all training tuples of the form $(s,r,o,T)$ are reduced to $(s,r,o,t)$, i.e., with a singleton time interval, where $t$ = $t_b$, the start time of $T$. 
\shortname{} sets a fact recurrence score, $\phi^\text{Rec}$, as follows:

\begin{enumerate}[leftmargin=*,partopsep=0pt,topsep=0pt] \raggedright
\item If $(s,r,o, \star) \notin \KBtrain$, 
set $\phi^\text{Rec} = 0$.
\item Else, if $r$ is not recurrent, set $\phi^\text{Rec}= b_r$. This allows the model to learn to penalize repetition of relations that do not recur.
\item Find time gap ($\delta$) to its closest recurrence:
\begin{align} \label{eq:minDiffDelta}
\delta &= \min_{\{(s,r,o,t')\in\KBtrain:\,t'\ne t\}}|t-t'|.
\end{align}
Then, set 
\begin{multline}
\phi^\text{Rec}(s,r,o,T=[t_b,t_e]) = \\
\phi^\text{Rec}(s,r,o,t_b) =
w_r \Normal\bigl(\delta|\mu_r,\sigma_r\bigr) + b_r.
\end{multline}
\end{enumerate}

The intuition is that $\phi^\text{Rec}$ should penalize the proposed $(s,r,o,T)$ if $\delta$ is not close to the mean gap~$\mu_r$. For example, $($Presidential election, held in, USA, 2017$)$ should be penalized, if $($Presidential election, held in, USA, 2016$)$ is known, and the event reoccurs every 4 years ($\mu_r=4, \sigma_r\approx0$).

\paragraph{Relation Pairs Score:}
\shortname{} also learns soft time constraints between pairs of relations. We describe this mechanism for subjects; objects are handled analogously. For each relation pair $(r,r')$, we maintain four parameters, $\mu_{rr'}$, $\sigma_{rr'}$, $b_{rr'}$ and $w_{rr'}$, whose purpose we will describe presently.
As with recurrence scores, all training tuples $(s,r,o,T)$ are reduced to $(s,r,o,t)$, where $t$ = $t_b$, the start time of $T$. Given the candidate tuple $(s,r,o,t)$ to score, we collect fact tuples
\begin{align}
\{f_i = (s,r_i,o_i,t_i) \in \KBtrain, r_i\ne r \},
\end{align}
$sc(f_i)=\Normal(t-t_i|\mu_{rr_i}, \sigma_{rr_i})+b_{rr_i}$
having the same subject but a different relation, into the set called~$\KBpair(s)$. The $i^{\text{th}}$ tuple in $\KBpair(s)$ is scored as $sc(f_i)=\Normal(t-t_i|\mu_{rr_i}, \sigma_{rr_i})+b_{rr_i}$. This represents the contribution of $f_i$ in the validity of candidate tuple, based on their (signed) time difference, and typical time differences observed between these two relations. $\phi^{\text{Pair}}_\text{sub}$ needs to aggregate these over~$f_i$. The (trained) parameter $w_{rr'}$ measures how much the times associated with $r'$ influence our belief in $(s,r,o,t)$. Using these, we define the weighted average
\begin{align*}
\phi^{\text{Pair}}_\text{sub}(s,r,o,t) &= \!\!
\sum_{f_i\in \KBpair(s)} \!\!
sc(f_i) \frac{\exp(w_{rr_i})}{\sum_{f_j} \exp(w_{rr_j})}.
\end{align*}
A similar $\phi^{\text{Pair}}_\text{obj}$ score is computed for the object entity, and overall $\phi^{\text{Pair}}=\phi^{\text{Pair}}_\text{sub}+\phi^{\text{Pair}}_\text{obj}$ (See Figure~\ref{fig:block-dig}\,(b)).
The \textbf{final scoring function} of \shortname{} is  \begin{multline}
\!\! \phi(s,r,o,T) = \phi^{\base{}}(s,r,o,T)  \\
+ \kappa \phi^\text{Pair}(s,r,o,T) +
\lambda \phi^\text{Rec}(s,r,o,T),
\label{eq:TimePlex-all}
\end{multline}
where $\kappa$ and $\lambda$ are model hyperparameters.

\subsection{Training}
\label{sec:TimePlex:Train}
We train \shortname{} in a curriculum of two phases. In the first phase, we optimize embeddings for all entities, relations and time-instants by minimizing the log-likelihood loss using only the base model~TX. We compute the probability of predicting a response $o$ for a query $(s, r, ?, T)$ as:
\begin{align}
\Pr(o|s,r,T) &= \frac{\exp(\phi^{\base{}}(s,r,o,T))}{\sum_{o'} \exp(\phi^{\base{}}(s,r,o',T))}
\end{align}
We can similarly compute $\Pr(s|r,o,T)$ and similar terms for time instant queries, e.g., $\Pr(o|s,r,t)$ and $\Pr(t|s,r,o)$. We then convert every $(s,r,o,T=[t_b,t_e])\in\KBtrain$ in time-instant format by enumerating all $(s,r,o,t)$, for $t\in[t_b,t_e]$. 
Training of embeddings minimizes the \textbf{log-likelihood loss}:
\begin{multline}
- \!\! \sum_{\langle s,r,o,t \rangle\in KB^{\train}} \!\! \Bigl( \log \Pr(o|s, r,t; \theta) \\[-2ex]
\\[-3ex] ~~~~~~~~~~~~ + \log \Pr(s|o,r,t; \theta)\\
\\[-3.5ex] + \log \Pr(t|s,r,o; \theta) \Bigr)
\label{eq:SoftMaxLoss}
\end{multline}
In the second phase, we freeze all embeddings and train the parameters of the recurrence and pairs models. Here, too, we use the log-likelihood loss, except that $\phi^{\base{}}$ is replaced by the overall $\phi$ function. Parameters $\mu_{rr'}$ and $\sigma_{rr'}$ of the relation-pairs model component are not trained via backpropagation. Instead, they are fitted separately, using the difference distributions for the pair of relations in the training KB. This improves the overall stability of training. 



\subsection{Inference}
\label{sec:TimePlex:Test}

At test time, for a link prediction query, \shortname{} ranks all entities in decreasing order of $\Pr(o|s,r,T)$ or $\Pr(s|r,o,T)$ scores. 
For time prediction, its goal is to output a predicted time 
duration $T^\pred$. We first compute a probability distribution over time instants $\Pr(t|s,r,o)=\frac{\exp(\phi(s,r,o,t))}{\sum_{t'\in{\Tall}}\exp(\phi(s,r,o,t'))}$. We then greedily coalesce time instants to output the best duration. 
For greedy coalescing, we tune a threshold parameter $\theta_r$ for each relation $r$ using the dev fold (such that shorter $\theta_r$ prefers short duration and vice versa). We then initialize the predicted interval $T^\pred$ as $\arg\!\max_t \Pr(t|s,r,o)$. Then, as long as total probability of the interval, i.e., $\sum_{t\in {T^\pred}} \Pr(t|s,r,o)$ is less than $\theta_r$, we extend $T^\pred$ with the instant to its left or right, whichever has a higher probability.

\section{Experiments}
\label{sec:Expt}

We investigate the following research questions. (1)~Does \shortname{} convincingly outperform the best time-agnostic and time-aware KBC systems on link prediction and time interval prediction tasks? (2)~Are recurrent and pairwise features helpful in the final performance? (3)~Are \shortname's time embeddings meaningful, i.e., do they capture the passage of time in an interpretable manner? (4)~Do \shortname{} predictions honor temporal constraints between relations?

\subsection{Datasets \& Experimental Setup}

\noindent{\bf Datasets: } 
We report on experiments with four standard TKBC datasets.
WIKIDATA12k and YAGO11k \citep{DasguptaRT2018HyTE} are two knowledge graphs with a time interval associated with each triple. These contain relational facts like (David Beckham, plays for, Manchester United; [1992, 2003]). ICEWS14 and ICEWS05-15 \citep{Garcia+2018TADM} are two event-based temporal knowledge graphs, with facts from Integrated Crisis Early Warning System repository. These primarily include political events with timestamps (no nontrivial intervals). We consider the time granularity for interval datasets as 1 year, and for ICEWS datasets as 1 day. 
See Table~\ref{table:dataset-stats} in Appendix \ref{sec:app:DatasetStatistics} 
for salient statistics of these datasets. By experimenting across the spectrum, from `point' events to facts with duration, we wish to ensure the robustness of our observations.\\

\citet{Garcia+2018TADM} also report performance on the Yago15k dataset. However, for this dataset, only 17\% of the facts have associated temporal information. In contrast, all the datasets we used had at least 99\% of facts with temporal information. Hence, we believe a temporal model will not substantially improve the performance of a time-agnostic model on this dataset. Note that TNTComplex \citep{lacroix2020tensor} also obtained only a slight improvement over a time-agnostic model on Yago15k, supporting our hypothesis. A contemporaneous work by \citep{git-tkbidata} proposed new multi-relational temporal Knowledge Graph based on the daily interactions between artifacts in GitHub. We leave exploration of this dataset for future work. 

\newcommand{\rr}[1]{\rotatebox[origin=c]{90}{#1}}

\begin{table*}
\centering
\begin{adjustbox}{max width=\textwidth}
\begin{small}
\begin{tabular}{|l|c|c|c|c|c|c|c|c|c|c|c|c|} \hline
Dataset$\to$ &
\multicolumn{3}{c|}{WIKIDATA12k} &
\multicolumn{3}{c|}{YAGO11k} &
\multicolumn{3}{c|}{ICEWS05-15} &
\multicolumn{3}{c|}{ICEWS14} \\ \hline
$\downarrow$Methods
        & \rr{MRR} & \rr{HITS@1} & \rr{HITS@10}
        & \rr{MRR} & \rr{HITS@1} & \rr{HITS@10}
        & \rr{MRR} & \rr{HITS@1} & \rr{HITS@10}
        & \rr{MRR} & \rr{HITS@1} & \rr{HITS@10} \\ \hline
\rowcolor{yellow!7}
\complex &   24.82 &    14.30    &  48.90
        &  18.14  &    11.46    &   31.11
        & 48.68   &    37.00    &  72.63
        &  45.50  & 33.87       &    69.73  \\
TA (\complex) &  22.78  & 12.69       &   46.00
            &  15.24   & 9.36        &  26.26
            & 49.23   & 37.6       &    72.69 
            & 40.97   & 29.58       &    63.87 \\
HyTE (TransE)    & 25.28  &    14.70    &  48.26
        &  13.55 &    3.32 &  29.81 
        & 23.73   &    3.11    &   62.76
        & 24.91 & 2.98 & 65.30 \\
DE-SimplE   & 25.29 & 14.68 & 49.05
            & 15.12 & 8.75 & 26.74
            & 51.30 & 39.20 & 74.80
            & 52.60 & 41.80 & 72.50 \\
TNT-Complex    & 30.10  &    19.73    &  50.69 
        &   18.01 &    11.02 &  31.28
        & 60.58   &    51.14    &   78.50
        & 56.72 & 47.04 & 75.40 \\
\rowcolor{green!4}
\shortname{} (base) &  32.38    &  22.03 &  52.79
 &   18.35  & 10.99 &  31.86
 &   63.91 &  \bf 54.62 &  81.42 
 &  60.25 &  51.29 &  77.05 \\
\rowcolor{green!8}
\shortname  & \bf 33.35  & \bf 22.78    &  \bf  53.20
            & \bf 23.64  & \bf 16.92 & \bf 36.71
            & \bf 63.99  & 54.51  & \bf 81.81
            & \bf 60.40 & \bf 51.50 & \bf 77.11 \\ \hline

\end{tabular}
\end{small}
\end{adjustbox}
\caption{Link prediction performance across four datasets. The last row reports results for \shortname(base) augmented with pair/recurrent features. 
}
\label{table:all-results}
\end{table*}

\paragraph{Algorithms compared:} We compare against our reimplementations of \complex{}, HyTE, TA-family, and TNT-Complex. In all cases we verify that our implementations give comparable or better scores as reported in literature. We combine HyTE and TA, with scoring functions from TransE, DistMult and CX and present the best results. We also compare against reported results in DE-SimplE.

\begin{table}[t]
\centering
\begin{adjustbox}{max width=\textwidth}
\begin{small}
\begin{tabular}{|l|c|c|} \hline
Datasets$\to$ & YAGO11k &
WIKIDATA12k \\ \hline
$\downarrow$Methods
& aeIOU  & aeIOU \\ \hline
\rowcolor{yellow!7}
HyTE & 5.41  & 5.41  \\

TNT-Complex & 8.40 & 23.35  \\
\rowcolor{green!4}
\shortname{} (base) &  14.21  &  26.20  \\ 
\rowcolor{green!8}
\shortname  & \bf 20.03 & \bf 26.36   \\ \hline

\end{tabular}
\end{small}
\end{adjustbox}
\caption{Time prediction performance.}
\label{table:timepred}
\end{table}

\paragraph{Experimental Details:} For all models, we optimize parameters with AdaGrad running for 500 epochs for all losses, with early stopping on dev fold. We control for an approximately comparable number of parameters and set dimensionality of 200 for all complex embeddings and 400 for all real embeddings. We follow other best practices in the literature, such as 
L2 regularization only on embeddings used in the current batch \cite{TrouillonWRGB2016Complex}, 
adding inverted facts $(o, r^{-1}, s, T)$ , using 1vsAll negative sampling 
\cite{dettmers2017convolutional} whenever applicable, and using temporal smoothing for ICEWS datasets \cite{lacroix2020tensor}. 

Some instances in interval datasets have $t_b$ or $t_e$ missing. Following \citet{DasguptaRT2018HyTE}, we replace missing values by $-\infty$ or $+\infty$, respectively. For time prediction queries, we remove such instances from test sets. For ICEWS datasets we set $t_b=t_e$.  For time interval prediction, all models use our greedy coalescing inference from Section~\ref{sec:TimePlex:Test}.

For \shortname{}, we perform a grid search for all hyperparameters, 
and pick the best values based on MRR scores on valiations set. 
Hyperparameters for all datasets are described in Appendix \ref{sec:app:Hyperparameters}.
{\color{red}}

\subsection{Results and Observations}

Table~\ref{table:all-results} compares all algorithms for link prediction. We find that the best performing baseline among existing TKBC systems is the recently proposed TNT-Complex model. \shortname{} outperforms TNT-Complex by over 3 MRR points in ICEWS datasets. Its gains (3.25 and 5.6 pts) are even more pronounced in interval datasets. 
All differences are statistically significant using paired t-test with $p<0.01$. These scores establish a new state of the art for link prediction on all four datasets.

A contemporaneous work, ATiSE \cite{nayyeri2020gaussiantemporal} models KB entities and relations using time dependent Gaussian embedding, but show weaker performance (see Table~\ref{table:all-results} and Table~\ref{table:full_filtering_vs_partial_filtering}).

We are the first to look at the task of predicting \textit{time intervals}, and we report performance using our novel aeIOU metric (Table~\ref{table:timepred}).
We see that \shortname{} outperforms TNT-Complex on both datasets, with a huge 11+ pt jump on the Yago11K dataset. It is also noteworthy that even the base model of \shortname{} is consistently better than TNT-Complex across all experiments. 

\textbf{On Pair/recurrent features:}
We find that recurrent features are very helpful in both interval datasets, and significantly improve link prediction performance. Relation pair
features particularly help in YAGO11k --- over 5 pt \aeiou{} boost in time prediction, but on WIKIDATA12k they make only a marginal difference.  On inspecting the datasets, we find that 78\% of entities in WIKIDATA12k are seen with a single, recurring relation (such as \textit{award received}, or \textit{member of sports team}); therefore, relation pair features cannot help.

ICEWS datasets are scraped from news events. On inspecting the datasets, we find that the events do not follow any temporal ordering and are fairly non-regular in event recurrence as well. Hence, \shortname's improvements over the base model are limited. We further investigate the differing performance on datasets and the value of pair features in the next section.

\subsection{Diagnostics}
\label{sec::Expt:Diagnostics}

\noindent{\bf Time gap vs.\ embedding distances: }
Longevity of relations, or gaps between events, are often determined by physical phenomena that are smooth and continuous in nature. Therefore, we expect the embedding of the year 1904 to be closer to that of 1905 compared to the embedding of, say, 1950.

To validate this hypothesis, we compute mean L2 distance between embeddings of time instants which are apart by a given time gap. To filter noise, we drop instant pairs with extreme gaps that have low support (less than 30). For WIKIDATA12k we used embeddings of years $[1984, 2020]$ and for YAGO11k we use embeddings of years $[1958,2017]$.

\begin{figure}[t]
\centering
\begin{tikzpicture}
    \tikzstyle{every node}=[font=\small]
    \begin{axis}[width=.55\hsize, height=.5\hsize,
      title style={yshift=-1.5ex,},
      ylabel style={yshift=-1ex},
      xlabel style={yshift=1.5ex},
      title=WikiData12k, ylabel=Distance$\rightarrow$,
      xlabel=$\Delta\text{years}\rightarrow$,
      legend pos=outer north east, legend columns=1, 
      xmajorgrids=true, ymajorgrids=true, grid style=dashed]
    \addplot [only marks, mark=square*,
      mark options={fill=blue!40}]
    table [x index=0,y index=1] {images/WIKIDATA12k.csv};
\end{axis}
\end{tikzpicture}
\begin{tikzpicture}
    \tikzstyle{every node}=[font=\small]
    \begin{axis}[width=.55\hsize, height=.5\hsize,
      title style={yshift=-1.5ex,},
      ylabel style={yshift=-1ex},
      xlabel style={yshift=1.5ex},
      title=YAGO11k,
      xlabel=$\Delta\text{years}\rightarrow$,
      legend pos=outer north east, legend columns=1, 
      xmajorgrids=true, ymajorgrids=true, grid style=dashed]
    \addplot [only marks, mark=square*,
      mark options={fill=blue!40}]
    table [x index=0,y index=1] {images/YAGO11k.csv};
    \end{axis}
\end{tikzpicture}
\caption{$L_2$ distances (y-axis) between \shortname{} time embeddings increase with time gap (x-axis).}
\label{fig:time-duration-l2-cos}
\end{figure}
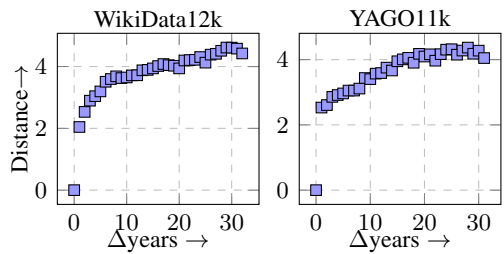

Figure~\ref{fig:time-duration-l2-cos} shows that $L_2$ distance between pairs of time embeddings increases with the actual year gap between them.
Since we enumerate all time points in the given fact time-interval, years that are closer share a lot of facts (triples), and are hence closer in the embedding space. This has a smoothing effect on time embeddings. Hence they correlate well with actual time-gaps. This strongly suggests that the time embeddings learnt by \shortname{} naturally represent physical time.

\paragraph{Temporal ordering of relation pairs:} 
Both YAGO11k and WIKIDATA12k contain relations with temporal dependencies, e.g., \emph{bornInPlace} should always precede \emph{diedInPlace} for the same person. We now study whether \shortname{} models are able to learn these natural constraints from data.


We first exhaustively extract all relation pairs $(r_1, r_2)$,
where the existence of both $(s,r_1,\star,t_1)$ and $(s,r_2,\star,t_2)$
is accompanied by $t_1 < t_2$ at least 99\% of the time, with a minimum support of~100 entities~$s$.\footnote{The list of such relation pairs is given in the Appendix \ref{sec:app:RelationOrdering}} We now verify whether \shortname{} honors $r_1$ before $r_2$ when making predictions.

For each query $(?,r,o,t)$ in the test set, we check whether the top model prediction violates any known temporal ordering constraint in this list. For example, for a query \textit{(?, \textbf{hasWonPrize}, Nobel Prize, \textbf{1925})}, if the model predicted \textit{Barack Obama} and the KB already had Barack Obama born in Hawaii in 1961, then this will be considered as an ordering violation. 
Table~\ref{table:temporal-constraints-eval} reports the number such violations as fraction of test set size. \shortname{} significantly reduces such errors for YAGO11k; this is also reflected in its superior time prediction performance. For WIKIDATA12k, the errors for \shortname{} (base) are already low, hence pair features are not found to be particularly helpful.



\begin{table}
\resizebox{\hsize}{!}{
\begin{small}
\begin{tabular}{|c|c|c|}
\hline
\textbf{}         & \textbf{YAGO11k} & \textbf{WIKIDATA12k} \\ \hline
\textbf{CX}       & 10.04 & 0.7                           \\ 
\textbf{HyTE}     &  7.2 & 0.4                                \\ 
\textbf{TNT-Complex}     & 8.82 & 0.3                              \\ 
\textbf{\shortname{} (Base)} &  6.6 & 0.3                     \\ 
\textbf{\shortname} & \bf 1.9 & \bf 0.2                \\ \hline
\end{tabular}
\end{small}
}
\caption{Ordering constraint violations among top predictions of various models (\text{\%} of facts in test set).}
\label{table:temporal-constraints-eval}
\end{table}

\begin{figure}[t]
  \centering
  \includegraphics[width=0.9\columnwidth]{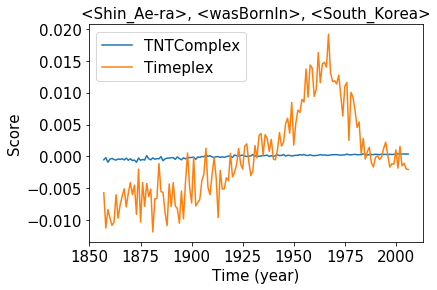}
  \caption{Time prediction comparison for two systems.}
\label{fig:scores_across_time}
\end{figure}

As an illustrative example, we consider the time prediction query \textit{(Shinae-ra, wasBornIn, South Korea, ?)}, with the gold answer 1969. The only other fact seen for \textit{Shinae-ra} in the train KB is \textit{(Shinae-ra, isMarriedTo, ChaIn-Pyo, (1995, -))}. 
\shortname{} predicts 1967 for this query (earning an aeIOU credit of 33.33). However, TNTComplex predicts 2013 (earning almost no credit) -- this also highlights that it does not capture commonsense that a person can marry only after they are born. 

To understand further, we plot the normalized scores for this query in time range $[1850,2010]$ in Figure \ref{fig:scores_across_time}. The peak around 1967 for the \shortname{} plot can be attributed to the fact that mean difference for \textit{isMarriedTo} and \textit{wasBornIn} relations is around 30 in the dataset. Standard tensor factorization models like TNT-Complex are unable to exploit this, but our Pair features provide a way to the model to make very reasonable predictions. Other similar plots can be found in the Appendix.

\section{Discussion} 
\shortname{} cannot exploit the influence that an entity can have on time difference distributions.  For example, the life expectancy of a person (mean difference between \textit{diedIn} and \textit{bornIn} events) would be around 85 in Japan, but 54 in Lesotho. Extending our model to learn separate parameters for each $\langle$rel, entity$\rangle$ pair may be difficult due to sparsity.  
Also, recurrent facts may admit exceptions: Winter Olympics are held every 4 years except for 1992 and 1994. However, we do not expect even humans to do well in such cases. Exceptions like these are sparse and difficult to learn, except by rote.


\section{Conclusion}
\label{sec:End}

We presented \shortname, a new TKBC framework, which combines representations of time with representations of entities and relations. It also learns soft temporal consistency constraints, which allow knowledge of one temporal fact to influence belief in another fact. \shortname{} exceeds the performance of existing TKBC systems. Time embeddings are temporally meaningful, and \shortname{} makes fewer temporal consistency and ordering mistakes. We also argue that current evaluation schemes for both link and time prediction have limitations, and propose more meaningful schemes.


\section*{Acknowledgements}
This work is partly supported by IBM AI Horizons Network grants.
IIT Delhi authors are supported by an IBM SUR award, grants by Google, Bloomberg and 1MG, Jai Gupta Chair professorship and a Visvesvaraya faculty award by the Govt.\ of India. The fourth author is supported by a Jagadish Bose Fellowship.  We thank IIT Delhi HPC facility for compute resources. We thank Sankalan, Vaibhav and Siddhant for their helpful comments on an early draft of the paper.

\bibliographystyle{acl_natbib}
\bibliography{tkbi,voila}

\clearpage
\twocolumn[\centering \bfseries \large \ztitle \\
(Appendix) \par\smallskip]
\appendix

\section{Dataset statistics}
\label{sec:app:DatasetStatistics}
See Table \ref{table:dataset-stats} for some salient statistics of the datasets we used for experiments. Yago11k and Wikidata12k are interval based datasets. ICEWS14 and ICEWS05-15 are instant based datasets. 

\begin{table*}[ht!]
\centering
\resizebox{0.6\hsize}{!}{
\begin{tabular}{|c|c|c|c|c|}
\hline
\textbf{}                       & \textbf{YAGO11k} & \textbf{WIKIDATA12k} & \textbf{ICEWS14} & \textbf{ICEWS05-15} \\ \hline
\textbf{Entities}               & 10622            & 12554                & 7128             & 10488               \\ \hline
\textbf{Relations}              & 10               & 24                   & 230              & 251                 \\ \hline
\textbf{\#Instants}             & 251              & 237                  & 365              & 4017             \\ \hline
\textbf{\#Intervals}            & 6651             & 2564                 & 0                & 0                   \\ \hline
\hline
\textbf{Train}                  & 16408            & 32497                & 72826            & 368962              \\ \hline
\textbf{Valid}                  & 2051             & 4062                 & 8941             & 46275               \\ \hline
\textbf{Test}                   & 2050             & 4062                 & 8943             & 46092               \\ \hline
\end{tabular}}
\caption{Details of datasets used.} 
\label{table:dataset-stats}
\end{table*}

\section{Discussion on time evaluation metrics}
\label{sec:app:DiscussionTimeMetrics}
We re-state the desired property P for a time evaluation metric-\\
Let $\vol(T^\eval \cap T^{\pred_1})=\vol(T^\eval \cap T^{\pred_2})$ $M(T^{\eval},T^{\pred_1})>M(T^{\eval},T^{\pred_2})$ if and only if $\vol(T^\eval \Cup T^{\pred_1})< \vol(T^\eval \Cup T^{\pred_2})$.\\

\noindent
\textbf{aeIOU satisfies P:}\\
For a fixed $\vol(T^\eval\cap T^\pred)$, we have $\aeiou(T^\eval,T^\pred)\propto 1/{\vol(T^\eval\Cup T^\pred)}$ (see Eqn \ref{eq:aeiou}). Hence, aeIOU satisfies property P. \\



\noindent
\textbf{IoU and gIOU do not satisfy P:}\\
{\bf IoU:} This metric gives 0 score to a model, if model's predicted interval does not intersect with the gold, irrespective of the hull. Hence IoU do not satisfy property P.\\
\noindent
{\bf gIoU:} 
Let us look at the following example. Suppose gold interval is [2002,2005], and consider 2 predictions- [1999,2001] and [1900,2001]. For both predictions, $\vol((T^\eval\Cup T^\pred)\setminus
(T^\eval\cup T^\pred))$ is zero, so the hull for the two predictions will be ignored (see Eqn \ref{eq:gIOU}), resulting in same scores for both predictions. Hence gIoU does not satisfy property P.\\

\noindent
{\bf Model Performance with respect to various time evaluation metrics}:\\
Table \ref{table:timepred-all-scores} reports the TAC, gIOU, and IOU scores of various temporal methods discussed in the paper.
\begin{table*}[t]
\centering
\begin{adjustbox}{max width=1.3\textwidth}
\begin{small}
\begin{tabular}{|l|c|c|c|c|c|c|c|c|} \hline
Datasets$\to$ & \multicolumn{4}{c|}{\bf YAGO11k} &
\multicolumn{4}{c|}{\bf WIKIDATA12k} \\ \hline
$\downarrow$Methods
& TAC  & gIOU & IOU & aeIOU & TAC & gIOU & IOU & aeIOU \\ \hline
HyTE & 5.59 & 15.96  & 1.91 & 5.41 & 6.13  & 14.55 &  1.40 & 5.41 \\

TNT-Complex & 9.90 & 20.78 & 3.99 & 8.40 & 26.98 & 36.63 & 11.68 & 23.25 \\
\shortname{} (base) &  16.57  & 26.22 & 5.48 & 14.21 & 30.36 & 39.2 & \bf 13.20 & 26.20  \\ 
\shortname  & \bf 22.66  & \bf 32.64 &  \bf 8.24 & \bf 20.03 & \bf 30.71 & \bf 39.34 & 13.15 & \bf 26.36 \\ \hline

\end{tabular}
\end{small}
\end{adjustbox}
\caption{Time prediction performance using - TAC, gIOU, IOU and aeIOU}
\label{table:timepred-all-scores}
\end{table*}

\section{Temporal Constraints: Relation Ordering}
\label{sec:app:RelationOrdering}
Table \ref{sup:table:rel-order-yago} and \ref{sup:table:rel-order-wiki} lists automatically extracted high confidence relation orderings seen in Yago11k and Wikidata12k datasets respectively. These orderings are used to guide \sys{} at the time of training.
\begin{table}[ht!]
\centering
\scalebox{0.9}{
\begin{tabular}{|c|}
\hline
\textit{graduatedFrom $\xrightarrow{}$ diedIn}      \\ \hline
\textit{graduatedFrom $\xrightarrow{}$ hasWonPrize} \\ \hline
\textit{wasBornIn $\xrightarrow{}$ graduatedFrom}   \\ \hline
\textit{wasBornIn $\xrightarrow{}$ diedIn}          \\ \hline
\textit{wasBornIn $\xrightarrow{}$ isAffiliatedTo}  \\ \hline
\textit{wasBornIn $\xrightarrow{}$ hasWonPrize}     \\ \hline
\textit{wasBornIn $\xrightarrow{}$ playsFor}        \\ \hline
\textit{wasBornIn $\xrightarrow{}$ worksAt}         \\ \hline
\textit{wasBornIn $\xrightarrow{}$ isMarriedTo}     \\ \hline
\textit{isAffiliatedTo $\xrightarrow{}$ diedIn}     \\ \hline
\textit{worksAt $\xrightarrow{}$ diedIn}            \\ \hline
\textit{isMarriedTo $\xrightarrow{}$ diedIn}        \\ \hline
\end{tabular}}
\caption{High confidence (99\%) relation orderings extracted from YAGO11k.}
\label{sup:table:rel-order-yago}
\end{table}

\begin{table}[ht!]
\centering
\scalebox{0.9}{
\begin{tabular}{|c|}
\hline
\textit{educated at $\xrightarrow{}$ position held}                                       \\ \hline
\textit{educated at $\xrightarrow{}$ employer}                                            \\ \hline
\textit{educated at $\xrightarrow{}$ member of}                                           \\ \hline
\textit{educated at $\xrightarrow{}$ award received}                                      \\ \hline
\textit{educated at $\xrightarrow{}$ academic degree}                                     \\ \hline
\textit{educated at $\xrightarrow{}$ nominated for}                                       \\ \hline
\textit{instance of $\xrightarrow{}$ head of government}                                  \\ \hline
\textit{residence $\xrightarrow{}$ award received}                                        \\ \hline
\textit{academic degree $\xrightarrow{}$ nominated for}                                   \\ \hline
\textit{spouse $\xrightarrow{}$ position held}                                            \\ \hline
\textit{located in the administrative}\\ 
\textit{territorial entity $\xrightarrow{}$ award received} \\ \hline
\end{tabular}}
\caption{High confidence (99\%) relation orderings extracted from WIKIDATA12k}
\label{sup:table:rel-order-wiki}
\end{table}

\section{Time prediction performance across relation classes}
\label{sec:app:TimePredAcrossRelations}
Instant relations include \textit{wasBornIn, diedIn, hasWonPrize}, which are events that don't span an interval.\\
Short relations include \textit{graduatedFrom, playsFor} whose duration averages less than 5 years.\\
Long relations include \textit{isMarriedTo, isAffiliatedTo} whose duration averages more than 5 years.\\

\begin{table}[h]
\resizebox{\hsize}{!}{
\begin{small}
\begin{tabular}{|l|l|l|l|}
\hline
     & {\bf Instant} & {\bf Short} & {\bf Long} \\ \hline
TNT-Complex & 4.24    & 16.34 & 3.73 \\ \hline
Timeplex    & \textbf{18.39}   & \textbf{20.63} & \textbf{24.8} \\ \hline
\end{tabular}
\end{small}
}
\caption{aeIOU@1 across relation classes on YAGO11k}
\label{table:time_pred_relation_classes}

\end{table}

\section{Comparison of filtering methods}
\label{sec:app:ComparisonOfFilteringMethods}
In Table \ref{table:full_filtering_vs_partial_filtering}, we report the performance of most competitive baseline and \shortname, the reported performance use a filtering strategy that does not enumerate time points in an interval and filters out entities on exact matching time-interval. Note that our model consistently outperforms TNT-Complex, even with a stricter filtering. 

\begin{table*}[ht!]
\centering
\resizebox{0.9\textwidth}{!}{
\begin{tabular}{|c|c|c|c|c|c|c|c|}
\hline
            & \textbf{Learning Rate} & \textbf{Reg wt} & \textbf{Batch size} & \textbf{Temporal smoothing} & \textbf{$\alpha$} & \textbf{$\beta$} & \textbf{$\gamma$} \\ \hline
YAGO11k     & 0.1                    & 0.03            & 1500                & 0.0                           & 5.0            & 5.0           & 0.0            \\ \hline
WIKIDATA12k & 0.1                    & 0.005           & 1500                & 0.0                           & 5.0            & 5.0           & 5.0            \\ \hline
ICEWS05-15  & 0.1                    & 0.005           & 1000                & 5.0                         & 5.0            & 5.0           & 5.0            \\ \hline
ICEWS14     & 0.1                    & 0.005           & 1000                & 1.0                        & 5.0            & 5.0           & 5.0            \\ \hline
\end{tabular}}
\caption{Hyperparameters for training \shortname(base) model embeddings on various datasets, tuned on MRR for validation set. Temporal smoothing was found to help on ICEWS datasets, however it gave no improvement for interval datasets. We tuned the parameters in a staged manner - first we tune learning rate ($lr$), regularization weight ($r$), batch size($b$), and temporal smoothing weight ($ts$). We performed a random search in the following ranges: $lr$ $\in$ $[0.0001, 1.0]$, $r$ $\in$ $[0.0001, 1.0]$, $b$ $\in$ $[100, 5000]$, and $ts$ $\in$ $[0.0001,10.0]$. The models were most sensitive to regularization weight and learning rate. After finding best values for these parameters, we tuned $\alpha$, $\beta$ and $\gamma$ weights for each dataset, doing a grid search over the set \{0.0, 2.0, 5.0, 7.0, 10.0\} }.
\label{sup:table:hyperparams_emb}
\end{table*}

\begin{table*}[ht!]
\centering
\resizebox{1.1\columnwidth}{!}{
\begin{tabular}{|c|c|c|c|c|c|c|}
\hline
                       & \multicolumn{3}{c|}{\bf WIKIDATA12k} & \multicolumn{3}{c|}{\bf YAGO11k} \\ \hline
Method                 & \rr{MRR}     & \rr{HITS@1}   & \rr{HITS@10}   
                       & \rr{MRR}     & \rr{HITS@1}   & \rr{HITS@10} \\ \hline

TNT-Complex & 27.35  & 17.59 & 48.51
                      & 15.78  & 10.21 & 28.64 \\ \hline

\shortname{}           & \bf 30.61         & \bf 20.79          &  \bf 51.78           
                       & \bf 22.77       & \bf 16.33          &  \bf 36.3 \\ \hline
\end{tabular}}

\caption{Performance of the best models using a filtering strategy that does not enumerate time points in an interval, and filters on an exact match instead. We find that while \shortname{} convincingly outperforms the previous SOTA TNT-Complex using this filtering strategy as well.}
\label{table:full_filtering_vs_partial_filtering}
\end{table*}


\section{Ablation Study}
\label{sec:app:AblationStudy}
In this study, we remove each component of \shortname\ (see equation~\ref{eq:TimePlex-all}) by making either $\kappa$=0 or $\lambda$=0, to understand the importance of each component (see Table~\ref{table:ablation}).

\begin{table*}[ht!]
\centering
\resizebox{1.1\columnwidth}{!}{
\begin{tabular}{|l|c|c|c|c|}
\hline
Prediction task$\to$       & \multicolumn{3}{c|}{\bf Link}              & \multicolumn{1}{l|}{\bf Time interval} \\ \hline
$\downarrow$Method                & \rr{MRR} & \rr{HITS@10} & \rr{HITS@1} & \rr{aeIOU@1} \\ \hline
\shortname              & 23.64        & 16.92            & 36.71           & 20.03            \\ \hline
\shortname -Pair        & 23.15        & 16.63            & 36.27           & 14.21            \\ \hline
\shortname - Rec        & 18.93        & 11.46            & 32.74           & 20.03            \\ \hline
\shortname - Pair - Rec & 18.35        & 10.99            & 31.86           & 14.21            \\ \hline
\end{tabular}}
\caption{Ablation study on Yago11k. Recurrence feature significantly help in link prediction while relation pair feature helps time-interval prediction.}
\label{table:ablation}
\end{table*}

\section{Details of Hyperparameters and Model training}
\label{sec:app:Hyperparameters}
All models are trained on a single NVIDIA Tesla K40 GPU. Our final model \shortname\ consist of a base model and two time-based gadgets.\\
{\it \shortname (base)} takes less than 10 minutes to train on all datasets except for ICEWS05-15, where it takes 80 minutes. Table \ref{sup:table:hyperparams_emb} lists best hyperparameters of \shortname(base) on respective dataset.\\
Both gadgets are trained independently in less than 10 minutes. The parameter $\lambda$=5.0 gave best results for interval datasets, while $\lambda$=1.0 gave best results on event datasets. On Yago11k $\kappa$=3.0, while for rest $\kappa$=0.0. The gadget weights are L2 regularized, with a regularization penalty of 0.002. The model use 100 negative samples per correct fact for training. 

\section{Model parameters}
\label{sec:app:ModelParameters}
The number of parameters for the \shortname and baseline models are compared in Table \ref{table:param_comparison}.
\begin{table*}[ht]
\centering
\resizebox{\columnwidth}{!}{
\begin{tabular}{|c|c|}
\hline
{\bf Models}         & {\bf Number of parameters} \\ \hline
HytE           & d($|E|+|T|+|R|$)                     \\ \hline
DE-SimplE      & 2d($(3\delta + (1-\delta))|E|+|R|$)                     \\ \hline
TNTComplex     & 2d($|E|+|T|+4|R|$)                     \\ \hline
Timeplex(base) & 2d($|E|+|T|+6|R|$)                     \\ \hline
Timeplex       & 2d($|E|+|T|+6|R|$) + 2($|R|^2$ + $|R|$)                      \\ \hline
\end{tabular}}
\caption{Number of parameters for each model. For HyTE we assume bucket size = 1 here. $\delta$ is the fraction of dimension to represent time in DA-SimplE model.}
\label{table:param_comparison}
\end{table*}

\section{Training details of \shortname, HyTE}
\label{sec:app:TrainingDetails}

Each dataset spans along a \textit{time range}, with a certain \textit{time granularity}, which can be year, month or day. \sys{} learns a time embedding for every point in this time range, discretized on the basis of the dataset's granularity (years for the interval datasets WIKIDATA12k and YAGO11k, and days for ICEWS datasets).
At training time, \sys{} looks at a single time point at a time - for this, we sample a time point uniformly at random from the query interval $[t_b, t_e]$ associated with the fact. In contrast, HyTE maps each time point to bin (heuristically determined), making the data granularity coarser, and learns representation of these bins. HyTE looks at time points in an interval as well, but enumerates each interval fact to produce a separate fact for each time point beforehand. \\
Our method of sampling is efficient as the data size is unchanged. It also ensures each fact is sampled uniformly, not hurting link prediction performance by oversampling of long duration facts. 

HyTE time prediction: HyTE can only predict a bin for the test fact. To convert predicted bins to years (or days), we take a mean of all years seen with the predicted bin and then do greedy coalescing to output time interval in years.

\begin{table*}
\centering
\resizebox{1.3\columnwidth}{!}{
\begin{tabular}{|c|c|}
\hline
\multicolumn{2}{|c|}{
\includegraphics[width=1.4\columnwidth]{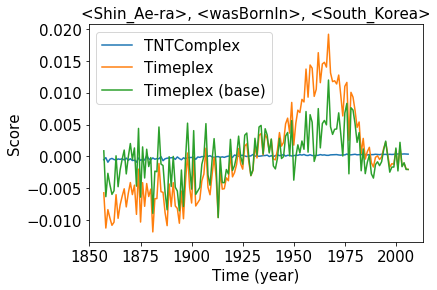}
} 
\\ \hline
\multicolumn{2}{|c|}{Info about query e1 in train set:} 
\\ \hline
\multicolumn{2}{|c|}{\textless{}Shin\_Ae-ra, isMarriedTo, Cha\_In-pyo\textgreater (1995, 3000)} 
\\ \hline
Gold answer & 1969 \\ \hline
Timeplex prediction & 1967 \\ \hline
Timeplex (base) prediction & 1967 \\ \hline
TNTComplex prediction & 2013
\\ \hline
\multicolumn{2}{|c|}{
\begin{minipage}{1.8\columnwidth}
(a) \shortname, \shortname (base) both predict the correct answer but\\ TNTComplex cannot model that one cannot marry before birth.
\end{minipage}
}
\\ \hline \hline
\multicolumn{2}{|c|}{
\includegraphics[width=1.4\columnwidth]{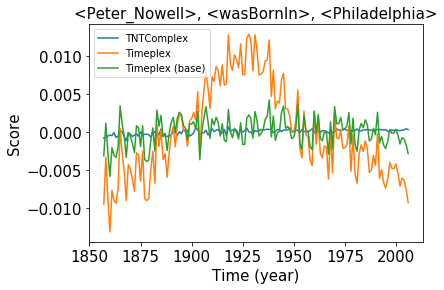}
}
\\ \hline
\multicolumn{2}{|c|}{Info about query e1 in train set:}
\\ \hline
\multicolumn{2}{|c|}{
\begin{minipage}{1.8\columnwidth}
\textless Peter\_Nowell, graduatedFrom, Wesleyan\_University\textgreater (1948, 3000)\\ \textless{}Peter\_Nowell, graduatedFrom, University\_of\_Pennsylvania\textgreater (1952, 3000)
\end{minipage}
}
\\ \hline
Gold answer & 1928 \\ \hline
Timeplex prediction & 1928 \\ \hline
Timeplex (base) prediction & 1938 \\ \hline
TNTComplex prediction & 1918
\\ \hline
\multicolumn{2}{|c|}{
\begin{minipage}{1.8\columnwidth}
(b) \shortname (base) cannot model that one is unlikely to graduate at the age of 10.\\
\shortname~(base) and TNTComplex do not have a clear vote like Timeplex.
\end{minipage}
} 
\\ \hline
\end{tabular}}
\caption{Comparing time prediction performance of \shortname, \shortname(base) and TNTComplex.}
\label{table:time_pred_appendix}
\end{table*}

\section{More diagnostics}
\label{sec:app:MoreDiagnostics}
We plot the normalized scores of \shortname, \shortname(base), and TNTComplex for different time queries in time range $[1850,2010]$ in \tablename~\ref{table:time_pred_appendix}. Figure (a) highlights how TNTComplex model fails to learn that one cannot marry before birth. Figure (b) shows how with the limited background knowledge on the subject in question, \shortname{} can predict the gold time-interval. 

\end{document}